# Giant Piezoresistive Effect and Strong Band Gap Tunability in Ultrathin InSe upon Biaxial Strain

*Qinghua Zhao, Tao Wang\*, Riccardo Frisenda\*, Andres Castellanos-Gomez\**

Q. Zhao, Prof. T. Wang
State Key Laboratory of Solidification Processing, Northwestern Polytechnical University, Xi'an, 710072, P. R. China
Key Laboratory of Radiation Detection Materials and Devices, Ministry of Industry and Information Technology, Xi'an, 710072, P. R. China
E-mail: taowang@nwpu.edu.cn

Q. Zhao, Dr. R. Frisenda, Dr. A. Castellanos-Gomez
Materials Science Factory. Instituto de Ciencia de Materiales de Madrid (ICMM-CSIC), Madrid, E-28049, Spain.

E-mail: riccardo.frisenda@csic.es; andres.castellanos@csic.es

**Abstract:** The ultrathin nature and dangling bonds free surface of two-dimensional (2D) semiconductors allow for significant modifications of their band gap through strain engineering. Here, thin InSe photodetector devices are biaxially stretched, finding, a strong band gap tunability upon strain. The applied biaxial strain is controlled through the substrate expansion upon temperature increase and the effective strain transfer from the substrate to the thin InSe is confirmed by Raman spectroscopy. The band gap change upon biaxial strain is determined through photoluminescence measurements, finding a gauge factor of up to ~200 meV/%. We further characterize the effect of biaxial strain on the electrical properties of the InSe devices. In the dark state, a large increase of the current is observed upon applied strain which gives a piezoresistive gauge factor value of ~450-1000, ~5-12 times larger than that of other 2D materials and of state-of-the-art silicon strain gauges. Moreover, the biaxial strain tuning of the InSe band gap also translates in a strain-induced redshift of the spectral response of our InSe photodetectors with $\Delta E_{\text{cut-off}}$ ~173 meV at a rate of ~360 meV/% of strain, indicating a strong strain tunability of the spectral bandwidth of the photodetectors.

**Introduction**

Strain engineering, the modification of the optical, magnetic, electrical, and optoelectronic properties of a given material by applying an external mechanical deformation to its crystal lattice, is establishing itself as one of the most prospective strategies to controllably modify the properties of two-dimensional (2D) materials.[1] In fact, the lack of dangling bonds on their surface makes them extremely resilient to the mechanical deformation without fracture,[2] even approaching the theoretical limit (predicted by Griffith) for defect-free materials.[3] The capability of applying very large deformations together with strain sensitive band-structures makes of 2D materials a very suitable family of materials for strain engineering. Based on this outstanding stretchability and strain engineered band-structure, novel strain-tunable devices for information, sensor and energy-saving technologies, usually referred as straintronics,[4] have been recently reported. In fact, very recently flexible broadband photodetectors based on continuous strain modulation,[5] micro stress sensors,[6] atomic-thin nanogenerators based on piezotronics,[7] and spatially and spectrally isolated quantum emitters on a pre-patterned rigid substrate have been achieved.[8]

During the last years, multiple works studying the strain tunability of the band gap of several 2D semiconductors, including transition metal dichalcogenides (TMDCs), black phosphorus (bP) and other 2D semiconductors, have been reported (see **Table 1**).[5a, 6c, 9] Very recently, InSe has shown sizeable larger strain tunability with respect to transition metal chalcogenides (TMDCs) and black phosphorus upon uniaxial strain loading and local strain modification.[9ab, 9ac, 10] According to the works reported for TMDCs and bP, biaxial strain usually yields stronger band gap tunability than uniaxial strain because of the larger lattice deformation in both crystal orientations.[9ae] Although recent calculations also predicted that biaxial strain should have a stronger effect on the InSe band structure than uniaxial strain,[11] its experimental realization is still lacking.

Here we experimentally study the band gap modification in ultrathin InSe by biaxial strain. We fabricate InSe photodetectors onto polycarbonate (PC) allowing us to control the applied biaxial strain through the substrate expansion upon temperature increase. Through Raman spectroscopy we verify that biaxial strain is effectively transduced from the substrate expansion. Photoluminescence (PL) measurements are used to probe the effect of biaxial strain tuning of the InSe band gap. We found a strong thickness dependence of the strain-tunability of the band gap, reaching ~200 meV/% of biaxial strain for ultrathin (~5 layers) InSe flakes. With electrical transport measurements we found a large increase of the dark current upon biaxial straining giving a piezoresistive gauge factor of GF~450 to 1000, that can reach ~5-12 times larger than that of other 2D materials and of state-of-the-art silicon strain gauges (GF~200).[6, 12] Interestingly, biaxial strain also has a strong effect on the spectral response of our photodetector, redshifting the photocurrent spectra up to ~173 meV at a rate of ~360 meV/% of strain, indicating a very strong strain tunable spectral bandwidth.

**Results and discussions**

The Au-InSe-Au devices are fabricated by mechanical exfoliation of bulk InSe single crystals grown by the Bridgman method (the characterizations of bulk crystals have been reported in our previous work) with Nitto SPV 244 tape.[13] The cleaved crystallites are then transferred onto a Gel-Film (Gel-Pak, WF 6.0 mil × 4) stamp. Quantitative optical microscopy is used to identify and select ultrathin InSe flakes on the Gel-Film stamp. Then the selected flake is deterministically placed bridging a pair of gold electrodes pre-patterned on a target polycarbonate (PC) substrate.[14] Subsequently, a larger thin h-BN flake (30-50 layers) is placed on the top of the active region in the device to provide a full insulating encapsulation to slow down the environmental induced degradation of InSe.[15] Note that all these fabrication steps are carried out under ambient conditions within 30 minutes. **Figure 1** shows the details of the fabrication of a Au-InSe-Au device on the PC substrate. **Figure 1a** shows the schematic (top panel) and optical images obtained with reflection (middle panel) and transmission (bottom panel) mode of a selected ultrathin ~ 20 nm InSe flake (shown by inset picture) deterministically transferred bridging two 50 nm Au/ 5 nm Ti electrodes pre-patterned on the surface of PC substrate. **Figure 1b** shows the geometry (top panel) and pictures (middle and bottom panels) of the final devices after top encapsulation with h-BN. We chose a PC substrate because of the combination of its high thermal expansion (to yield sizeable biaxial strain upon heating, $\alpha = 64 \cdot 10^{-6}$ °C$^{-1}$) and its high Young's modulus (to ensure a good strain transfer, $E = 2.5$ GPa).[5a, 9l] We also fabricated a set of InSe devices on SiO$_2$/Si substrates (see **Figure S1** in the Supporting Information) that have negligible thermal expansion coefficient ($\alpha < 1 \cdot 10^{-6}$ °C$^{-1}$).[16] This set of devices is used as control samples to determine the role of the intrinsic temperature increase, without biaxial strain, on the observed features. This allows for the disentanglement of the temperature and strain effects on the observed features during the measurements. To passivate the defects existing in thin InSe flakes, thanks to the air species trapped at the interfaces, and reach a long-term stable working state with fast photodetection operation (as shown in **Figure S2** and **S3** in Supporting Information),[17] all the Au-InSe-Au devices have been annealed *in situ* in air at ~100 °C for around 2 hours on the micro-heater mounted on probe station before carrying out the Raman spectroscopy and optoelectronic characterizations discussed in this work.[18] In **Figure S4** and **Figure S5** we show how the current flowing through the devices evolves under 530 nm global illumination during the annealing process both on PC and SiO$_2$/Si substrates, as expected for a defects passivation process in InSe photodetector. Raman spectroscopy measurement in **Figure S6** indicates there is no structural change before and after annealing.

We first employ Raman spectroscopy to characterize the strain transfer from the PC substrate to the flake upon thermal expansion. **Figure 2a** shows Raman spectra acquired in the 20 nm thick InSe device at different PC substrate temperatures (from ~26 °C to ~100 °C), corresponding to a biaxial thermal expansion ranging from 0% up to 0.48%.[5a, 9l] We address the reader to the Supporting Information **Figure S7** for a second set of Raman measurements acquired on the same sample during another heating cycle to demonstrate the reproducibility of the thermal induced biaxial straining approach. Three Raman active in-plane modes $A_1'(1)$, $A_2''(1)$ and $A_2'(1)$ located at ~113 cm$^{-1}$, ~198 cm$^{-1}$ and ~226 cm$^{-1}$, and one out-of-plane $E''(2)$ located at ~176 cm$^{-1}$ are observed, which is consistent with hexagonal crystal structure of ultrathin InSe with $\varepsilon$ stacking sequence.[13, 17-19] All the Raman peaks shift towards lower Raman shifts upon thermal expansion, similar to recently reported experimental works on uniaxial strained InSe due to phonon softening.[9ac, 10a, 20] That is the increase of the covalent bonds length introduced by the applied tensile strain results in a weaker restoring force of vibrations, and thus lower phonon frequencies. As a control experiment we repeat the same measurements on an InSe device fabricated on SiO$_2$/Si (with negligible thermal expansion). In this control sample the Raman peaks position shift at much lower rate upon SiO$_2$/Si substrate temperature increase (see **Figure 2b**), indicating that the shift observed in the PC based device can be mostly attributed to the effect of biaxial strain. By subtracting the shift obtained on the SiO$_2$/Si substrate to that of the PC substrate, in **Figure 2c** we can determine the redshift rate due to biaxial strain: -1.48 cm$^{-1}$/%, -4.84 cm$^{-1}$/%, -5.32 cm$^{-1}$/% and -5.77 cm$^{-1}$/% of biaxial strain for the $A_1'(1)$, $E''(2)$, $A_2''(1)$ and $A_2'(1)$ Raman modes, respectively. We address the reader to the **Figure S8** in Supporting Information for another dataset acquired on a 13 nm thick InSe flake showing very similar Raman peak shift upon straining, demonstrating that a similar strain transfer is achieved for the 20 nm and the 13 nm InSe flakes. We attribute this good strain transfer, even for relatively thick InSe flakes, to the low Young's modulus of InSe ($E = 23 \pm 5$ GPa,[13] 10-20 times smaller than that of transition metal dichalcogenides, TMDCs)[21] as strain transfer from the substrate to the flake is inversely proportional to the Young's modulus of the flake. Indeed, finite element calculations predict a strain transfer of ~100% for InSe on PC substrates.[13] In order to probe if the top h-BN encapsulation has any effect on the strain transfer we have performed a control straining experiment on a ~13 layers InSe flake that has been partially encapsulated with h-BN (see **Figure S9** in the Supporting Information), finding very similar results on the unencapsulated and on the encapsulated parts.

The redshift rates of the Raman peaks are around 2 times larger than the value reported for uniaxial strained thin InSe.[9ab, 9ac, 10a] This information can be highly valuable as Raman spectroscopy is commonly used to monitor residual or built-in strains during the device fabrication and/or growth of other 2D materials. More interestingly, we can further calculate the Grüneisen parameters,[22] that describes the effect of a volume change on the vibrational properties, of the $A_1'(1)$, $E''(2)$, $A_2''(1)$ and $A_2'(1)$ Raman modes by using the obtained Raman mode shift rate, which take the values of 0.65, 1.38, 1.34 and 1.28. These values are comparable with those reported in the literature for uniaxial strained InSe.[9ac] The determination of the Grüneisen parameters through biaxial straining, however, has the advantage that (unlike in uniaxial strain) no assumptions about the Poisson's ratio value are needed.

We study the effect of the applied biaxial strain on the band gap of InSe through photoluminescence (PL). **Figure 3a** shows PL spectra acquired on InSe flakes transferred onto a PC substrate and onto a SiO$_2$/Si substrate at different temperatures. The PL spectra show a peak corresponding to the direct band gap transition at the Γ point of the Brillouin zone and it is thus a good probe of the band gap of InSe.[9ab, 23] Note that one can use the PL energy emission to determine the number of layers of InSe (see **Figure S10** of the Supporting Information). As for the Raman experiments, we use the measurements on the SiO$_2$/Si substrates as a control experiment to probe the intrinsic shift of the PL peaks upon temperature increase (without biaxial strain). This allows to determine the biaxial strain induced PL shift, subtracting the PL shift measured on SiO$_2$/Si substrates (only thermal contribution) to the PL shift measured on PC substrates (thermal + biaxial strain contribution). **Figure 3a** shows how the PL shift on PC substrates is much larger than that measured on SiO$_2$/Si substrates, indicating that biaxial strain strongly modifies the band gap of InSe. Interestingly, we have found a clear thickness dependence on the band gap strain tunability: thinner flakes are more sensitive to strain than thicker flakes. **Figure 3b** summarizes the PL shift rate measured for 19 InSe flakes (10 on PC and 9 on SiO$_2$/Si) 5 to 30 layers thick. By subtracting the two trends obtained for the PC and the SiO$_2$/Si substrates we can obtain the thickness dependent band gap gauge factor, i.e., the change of band gap per % of biaxial strain, of InSe that ranges from 195 ± 20 meV/% (for 5 layers thick InSe) to 63 ± 6 meV/% (for 30 layers thick InSe). This value is among one of the largest reported values for 2D semiconducting materials so far, as shown in **Table 1** and **Figure 3d**. Interestingly, for ultrathin flakes our band gap gauge factor is nearly twice that of uniaxial strained InSe,[9ab, 9ac] in very good agreement with recent DFT predictions.[11a]

We further study the effect of biaxial strain on the electronic properties on the Au-InSe-Au photodetector devices in the dark state. The details of basic optoelectronic characterizations of the annealed Au-InSe-Au device on PC and on $SiO_2$/Si substrates are shown in **Figure S2** and **Figure S3**, respectively. **Figure 4a** shows the current *vs*. voltage characteristics (*I-V*s hereafter) in linear scale as a function of the PC substrate temperature from ~23 °C to ~ 100 °C, leading to a biaxial strain in the 0% - 0.48% range.[5a] A significant increase in the slope of *I-V*s with temperature increase is observed, indicating an increase of conductivity of the device upon substrate thermal expansion (in good agreement with the observed band gap reduction under biaxial tension). The data is also plotted in semi-logarithmic scale (with current absolute value) to facilitate the quantitative comparison between different datasets. The current at $V = $ -1 V increases dramatically: from -0.6 pA at ~23 °C (0% strain) to -0.33 nA at 100 °C (0.48% strain), see inset in **Figure 4a**. In order to estimate the intrinsic contribution of the temperature increase (without strain) on the observed current change we repeat the measurement on a control device fabricated on a $SiO_2$/Si substrate (with very small thermal expansion) finding a negligible current change (**Figure 4b** and inset). A minor increase of the current value from -0.46 pA to -2.6 pA at -1 V, due to the increase of thermal excited carriers, is observed.[24] We thus attribute the observed current change in the PC device to a piezoresistive response of InSe to biaxial strain. In order to quantify this piezoresistive response, and to compare it with that of other materials, in **Figure 4c**, we extract the current absolute value flowing through the devices at 1 V and -1 V both on PC and SiO$_2$/Si substrate as a function of temperature. The calculated electrical gauge factor in our device, $GF = (I - I_0)/\varepsilon I_0$, reaches values of ~450 at 1 V and ~1076 at -1 V, ~5-12 times larger than that found for InSe under uniaxial strain loading and other strained 2D materials. In fact, GF values of ~150, ~220 and ~40 have been reported for single-, bi- and tri-layer MoS$_2$, respectively,[6c] upon biaxial strain and GF ~15-30 for tri-layer MoS$_2$ upon uniaxial strain.[25] For graphene sensors GF values up to ~ 125 has been reported.[6, 9ac, 26] Moreover, the large gauge factor and mechanical resilience of 2D InSe makes it even more suitable as a biaxial strain sensor than state-of-the-art silicon strain sensors (GF ~ 200) with a fracture strain of only ~0.7%.[12, 27]

The strong PL shift upon biaxial straining indicates that biaxial strain could be an efficient strategy to tune the spectral bandwidth of InSe based photodetectors. In order to study this possibility we measure the photocurrent of the InSe photodetector upon illumination with different wavelengths at a fixed bias of 1 V and power density of 35.4 mW/cm$^2$. We address the reader to the Materials and Methods section for details about the measurement configuration. **Figure 5a** shows the photocurrent spectra measured at different temperatures between ~23 °C and ~100 °C (corresponding to a biaxial strain range of 0% to 0.48%) in the InSe photodetector fabricated on PC.[5a, 9l] The overall spectra redshift upon biaxial strain, as expected from the strain-induced reduction of the band gap observed in the PL measurements. This can be seen more clearly in the Tauc plot representation (see inset) that allows for an estimation of the energy cut-off ($\Delta E_{\text{cut-off}}$) of the photodetector (the minimum detectable photon energy). The theoretical foundation of this technique (Tauc plot extrapolation) is based on the energy dependence of the above-band-gap absorption, which appears either as a square relation (direct-allowed-transition dominant) or a square-rooted relation (indirect-allowed-transition dominated) and due to the direct band gap of thin InSe we take a squared plot relation.[23b, 28] In order to determine whether the observed band gap reduction is mainly caused by the biaxial strain induced by substrate expansion, and not by the temperature increase, we perform again another control measurement on an InSe device fabricated on a SiO$_2$/Si substrate. **Figure 5b** shows how in the control device (with negligible thermal expansion) the redshift of the spectra is less pronounced (another example is shown in **Figure S11** in Supporting Information). Therefore, we can reliably extract the biaxial strain tunability of the energy cut-off ($\Delta E_{\text{cut-off}}$) directly from the measurements on the InSe device fabricated on PC. **Figure 5c** summarizes the energy cut-off ($\Delta E_{\text{cut-off}}$) values extracted for the different biaxial strain levels showing a marked linear trend. The slope of this linear relationship is higher than the band gap gauge factor giving a value of ~ 360 meV/%. This could be due to the intrinsic higher uncertainty in the Tauc plot extrapolation method (given the reduced number of data points at the absorption edge part of the spectra).

**Conclusion**

In summary, we have studied the effect of biaxial strain on the vibrational, photoluminescence, electrical and optoelectronic properties of ultrathin InSe. We found a strong shift of the photoluminescence spectra upon biaxial strain with a gauge factor ranging from 195 meV/% for 5 thick layers InSe to 63 meV/% for 30 layers thick InSe. We also found a giant piezoresponse with an electrical gauge factor of ~ 1000 in the dark state. Interestingly, we also demonstrate how the strain tunable band gap can be exploited to tune the spectral response of InSe photodetectors. This work demonstrates the potential of InSe for future straintronic devices like optical modulators or photodetectors with a strain tunable spectral range.

**Materials and Methods**

**Sample fabrication.** Thin InSe and h-BN flakes were mechanically exfoliated out of high-quality bulk single crystals. InSe bulk crystals were grown by Bridgman method and h-BN single crystals were provided by HQ Graphene. During the mechanical exfoliation process, Scotch tape, Nitto tape (Nitto Denko® SPV 224) and a Gel-Film (Gel-Pak, WF 6.0 mil × 4) stamp were used as reported somewhere else.[14a, 14c] An optical microscope (Motic BA310 Met-T) in transmission mode has been used for flake inspection and to select thin InSe flakes. Selected flakes were deterministically transferred to bridge the 50 nm Au/ 5 nm Ti electrodes pre-patterned on the PC and 280 nm $SiO_2$/Si substrates. Then we transfer a h-BN flake onto the surface to realize the full encapsulated devices. The pre-patterned electrodes have been fabricated by e-beam evaporation of 5 nm Ti + 50 nm Au through a metal shadow-mask (E321 Osilla®). Note that all the channel length of Au-InSe-Au devices is ~30 μm and all the fabrication steps were carried out under ambient conditions.

**Raman spectroscopy and photoluminescence (PL) measurements.** Both the temperature-dependent Raman and photoluminescence characterization of InSe flakes on PC and on 280 nm $SiO_2$/Si substrates have been carried out with a confocal Raman microscopy system (MonoVista CRS+ from Spectroscopy & Imaging GmbH). The Raman and PL spectra have been recorded using a 532 nm excitation laser at the incident power of 196 μW and a 50x objective with the integration time of 20 s (Raman) and 6 s (PL), respectively.

**Electronic and optoelectronic characterization.** Au-InSe-Au devices are characterized in a homebuilt probe station mounted inside a high-vacuum chamber reported somewhere else.[29] The electrical measurements (*I-V*, *I-t*) were performed with a source-measure unit (Keithley® 2450). Fiber-coupled light emitting diodes (LEDD1B – T-Cube LED driver, Thorlabs®) with wavelength from 365 nm to 940 nm, were coupled to a multimode optical fiber and projected onto the sample surface by a zoom lens, creating a light spot on the sample with the diameter of 600 μm.

Note that all the temperature control from ~23 °C to ~100 °C was realized by using a resistance ceramic miniature-heater (10 mm x 10 mm) mounted on the sample stage connected to current source and a thermocouple to display the temperature.

**AFM measurements.** The thickness of thin InSe flakes was measured by an ezAFM (by Nanomagnetics) atomic force microscope operated in dynamic mode. The cantilever used is Tap190Al-G by BudgetSensors with force constant 40 $Nm^{-1}$ and resonance frequency 300 kHz.

## ACKNOWLEDGEMENTS

This project has received funding from the European Research Council (ERC) under the European Union's Horizon 2020 research and innovation programme (grant agreement n° 755655, ERC-StG 2017 project 2D-TOPSENSE) and the European Commision, under the Graphene Flagship (Core 3, grant number 881603). RF acknowledges support from the Spanish Ministry of Economy, Industry and Competitiveness through a Juan de la Cierva-formación fellowship (2017 FJCI-2017-32919). QHZ acknowledges the grant from China Scholarship Council (CSC) under No. 201706290035. TW acknowledges support from the National Natural Science Foundation of China: 51672216. This work is also sponsored by Innovation Foundation for Doctor Dissertation of Northwestern Polytechnical University.

## COMPETING INTERESTS

The authors declare no competing financial interests.

## FUNDING

Spanish Ministry of Economy, Industry and Competitiveness: Juan de la Cierva-formación fellowship 2017 FJCI-2017-32919 EU H2020 European Research Council (ERC): ERC-StG 2017 755655 EU Graphene Flagship: Grant Graphene Core 3, 881603 National Natural Science Foundation of China: 51672216.

**TABLE 1 Summarized comparison of band gap tunability of two-dimensional semiconductors under strain engineering.** The $\varepsilon^{max}$ refers to the maximum strain applied on the flexible substrate or directly on a suspended 2D flake or induced by a pre-patterned substrate. (D) and (I) indicate direct and indirect band gap, respectively. * Data obtained in a polyvynilacetate (PVA) encapsulated sample. ** Data obtained based on CVD grown 2D materials.

| Materials | Strain type | Method/substrate | $\Delta E_g^{max}$, $\varepsilon^{max}$ | Gauge factor (meV/%) | Ref. |
|---|---|---|---|---|---|
| 1L MoS$_2$ | Uniaxial | Mechanical bending, PC | -81 meV, 1.8% | -45 ± 7 | [9a] |
| | | Mechanical bending, PMMA | -33 meV, 0.52% | -64 ± 5 | [9b] |
| | | Mechanical bending, PET | -38.4 meV, 0.8% | -48 | [9c] |
| | | Mechanical bending, PC | -57.5 meV, 1.37% | -42 | [9d] |
| | | Mechanical bending, PC | -44.5 meV, 1.06% | -42 | [9e] |
| | | MEMS mechanics, suspended | -49.4 meV, 1.3% | -38 ± 1** | [9f] |
| | | Mechanical bending, PVA | -300 meV, 1.7% | -125*, -61 | [9g] |
| | | Mechanical bending, PET | -36 meV, 0.64% | -56 | [9h] |
| | | Substrate stretching, PDMS | -15 meV, 4.8% | -3** | [9i] |
| | | Mechanical bending, PI | -31 meV, 0.4% | -78 ± 4 | [9j] |
| | Biaxial | Thermal expansion, PC | -65 meV, 0.48% | -135 | [5a] |
| | | Pressurized membranes, suspended | -500 meV, 5% | -99 | [9k] |
| | | Thermal expansion, PDMS, PP | -12.5 meV, -51.1 meV, 1% | -12.5, -51.1 | [9l] |
| | | Pre-patterned substrate, SiO$_2$ | -50 meV, 0.565% | -110 | [9n] |
| | | AFM indentation, suspended | --, 7% | -77.3±10 | [6c] |
| 2L MoS$_2$ | Uniaxial | Mechanical bending, PC | -32 meV$^{(D)}$, -77 meV$^{(I)}$, 0.6% | -53±10$^{(D)}$, -129 ± 20$^{(I)}$ | [9a] |
| | | Mechanical bending, PMMA | -25 meV, 0.52% | -48 ± 5 | [9b] |
| | | Mechanical bending, PET | -36.8 meV$^{(D)}$, -68.8 meV$^{(I)}$, 0.8% | -46$^{(D)}$, -86$^{(I)}$ | [9c] |
| | | Mechanical bending, PC | -78 meV, 1.6% | -49 ± 1 | [9ad] |
| | | Mechanical bending, PI | -12 meV$^{(D)}$, 56 meV$^{(I)}$, 0.36% | -34 ± 3$^{(D)}$, -155 ± 11$^{(I)}$ | [9j] |
| | Biaxial | AFM indentation, suspended | --, 7% | -116.7±10 | [6c] |
| 3L MoS$_2$ | Uniaxial | Pre-strained substrate, Gel-Film | -90 meV, 2.5% | -36 | [9o] |
| | Biaxial | piezoelectric substrate, PMN-PT | -60 meV, 0.2% | -300 | [9m] |
| | | AFM indentation, suspended | --, 7% | -22.7±6 | [6c] |
| 1L MoSe$_2$ | Uniaxial | Mechanical bending, PC | -40.7 meV, 1.07% | -38 ± 2 | [9d] |
| | | Mechanical bending, PC | -30 meV, 1.1% | -27 ± 2 | [9p] |
| | | Mechanical bending, PEN | -28 meV, 0.5% | -54.8 ± 5.8 | [9q] |
| | Biaxial | Thermal expansion, PP | -33 meV, 1% | -33 | [9l] |
| 1L WS$_2$ | Uniaxial | Mechanical bending, PC | -69 meV, 1.26% | -55 ± 2 | [9d] |
| | | Mechanical bending, PVA | -253 meV, 5.68% | -43** | [9g] |
| | | Mechanical bending, PET | -27.5 meV, 0.64% | -43 | [9h] |
| | | Mechanical bending, PEN | -31 meV, 0.5% | -61.2 ± 3.8 | [9q] |
| | | Mechanical bending, PET | -44 meV$^{(D)}$, -76 meV$^{(I)}$, 4% | -11**$^{(D)}$, -19**$^{(I)}$ | [9r] |
| | | Substrate stretching, PDMS | -20 meV, 16% | -1.3** | [9s] |
| | Biaxial | Thermal expansion, PP | -95 meV, 1% | -95 | [9l] |
| 1L WSe$_2$ | Uniaxial | Mechanical bending, PC | -72.5 meV, 1.48% | -49 ± 2 | [9d] |
| | | Mechanical bending, PVA | -176 meV, 1.7%<br>-137 meV, 2.56% | -109<br>-53** | [9g] |
|  |  | Mechanical bending, PEN | -20 meV, 0.35% | -53 ± 3.1 | [9q] |
|  |  | Mechanical bending, PC | -75.5 meV, 1.4% | -54 | [9t] |
|  |  | Mechanical bending, PETG | 101 meV, 2.1% | -48 | [9u] |
|  | Biaxial | Thermal expansion, PP | -63 meV, 1% | -63 | [9l] |
| 2L WSe$_2$ | Uniaxial | Mechanical bending, PETG | -68 meV, 1.51% | -45 | [9v] |
|  |  | Mechanical bending, PET | -45 meV$^{(D)}$, -40 meV$^{(I)}$, 2% | -22.5$^{(D)}$, 20$^{(I)}$ | [9w] |
|  |  | Mechanical bending, PETG | -110 meV, 2.1% | -52 | [9x] |
| 1L ReSe$_2$ | Uniaxial | Pre-strained substrate, Gel-Film | -70 meV, 1.64% | -43 | [9y] |
| 6L bP | Uniaxial | Mechanical bending, PET | 110 meV, 0.92% | 120 | [9z] |
| 18L bP | Uniaxial | Pre-strained substrate, Gel-Film | 700 meV, 5% | 100-140 | [9aa] |
| 6L bP | Uniaxial | Mechanical bending, PP | 132 meV, 1% | 132 | [9ae] |
|  | Biaxial | Thermal expansion, PP | 67 meV, 0.3% | 222 | |
| 4-8L InSe | Uniaxial | Mechanical bending, PP | -110 meV, 1.15% | -(90-100) | [9ab] |
| 4-35 nm InSe | Uniaxial | Mechanical bending, PET | -118 meV, 1.06% | -(80-150) | [9ac] |
| 5L InSe | Biaxial | Thermal expansion, PC | -26 meV, 0.13% | -200 | This work |

**FIGURES**

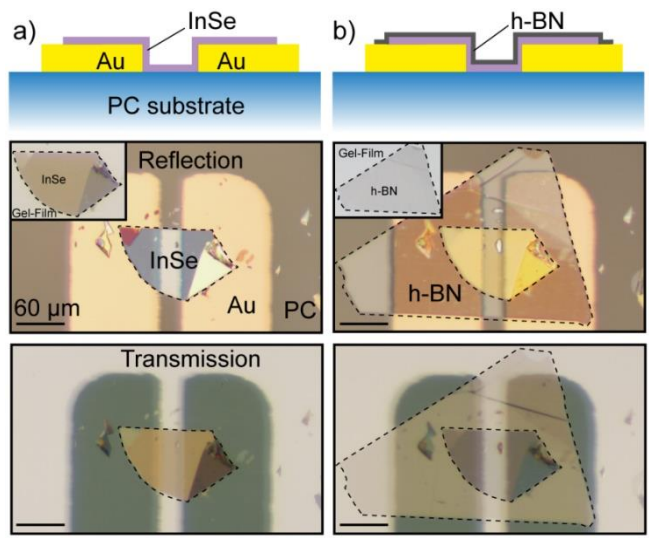

**Figure 1**. **Fabrication of Au-InSe-Au device on polycarbonate (PC) substrate.** a, b) Schematics (top) and optical microscopy images of a Au-InSe-Au device before (a) and after (b) encapsulating a h-BN flake on the surface with optical microscope reflection (middle) and transmission (bottom) mode. Insets: the optical microscopy images of selected thin InSe flake and h-BN fabricated on Gel-Film observed with transmission mode.

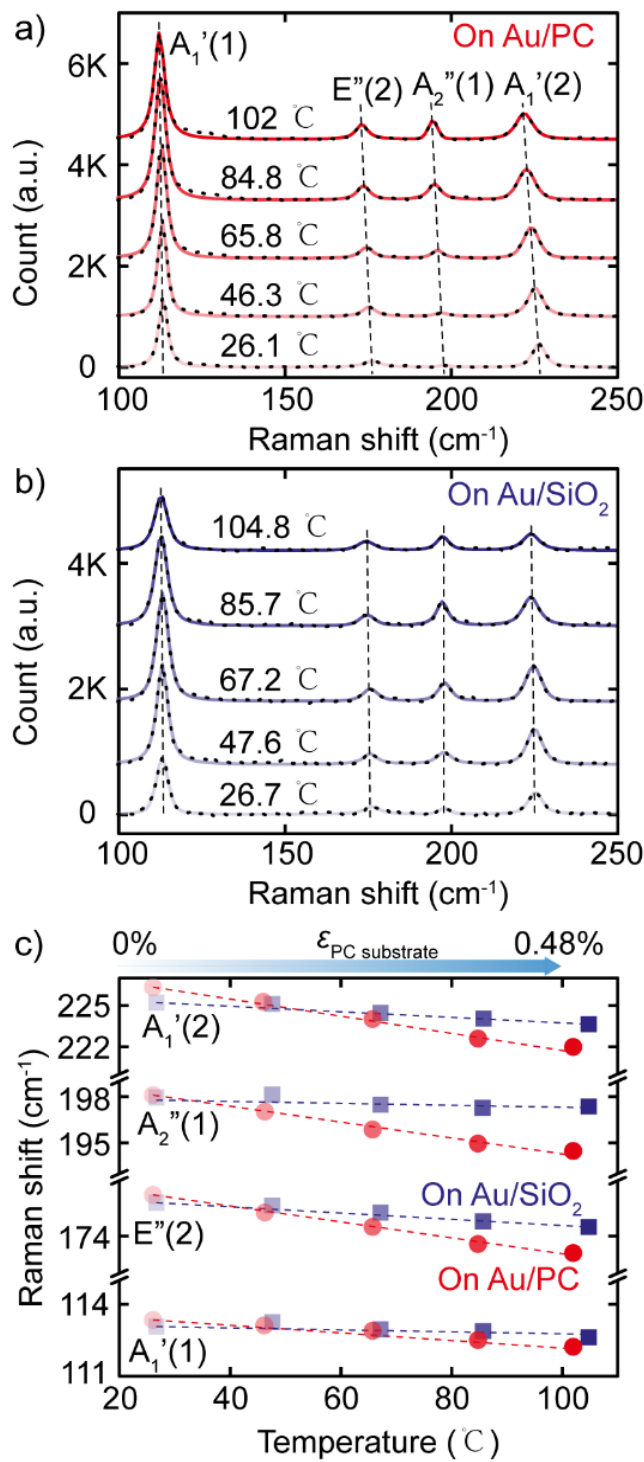

**Figure 2. Temperature-dependent Raman Spectroscopy of thin InSe flakes deposited on Au/PC and on Au/SiO$_2$/Si substrates.** a, b) Raman spectra of thin InSe recorded on Au/PC (a) and on Au/280 nm SiO$_2$/Si (b) substrates with 50× objective as a function of temperature (from ~26°C to ~ 100°C). (c) Temperature-dependency of four Raman active modes (A$_1$'(1), E''(2), A$_2$''(1) and A$_1$'(1)) of thin InSe on PC (red) and on Au/280 nm SiO$_2$/Si (blue) substrate.

The top axis in (c) indicates the biaxial strain induced by the thermal expansion of the PC substrate.

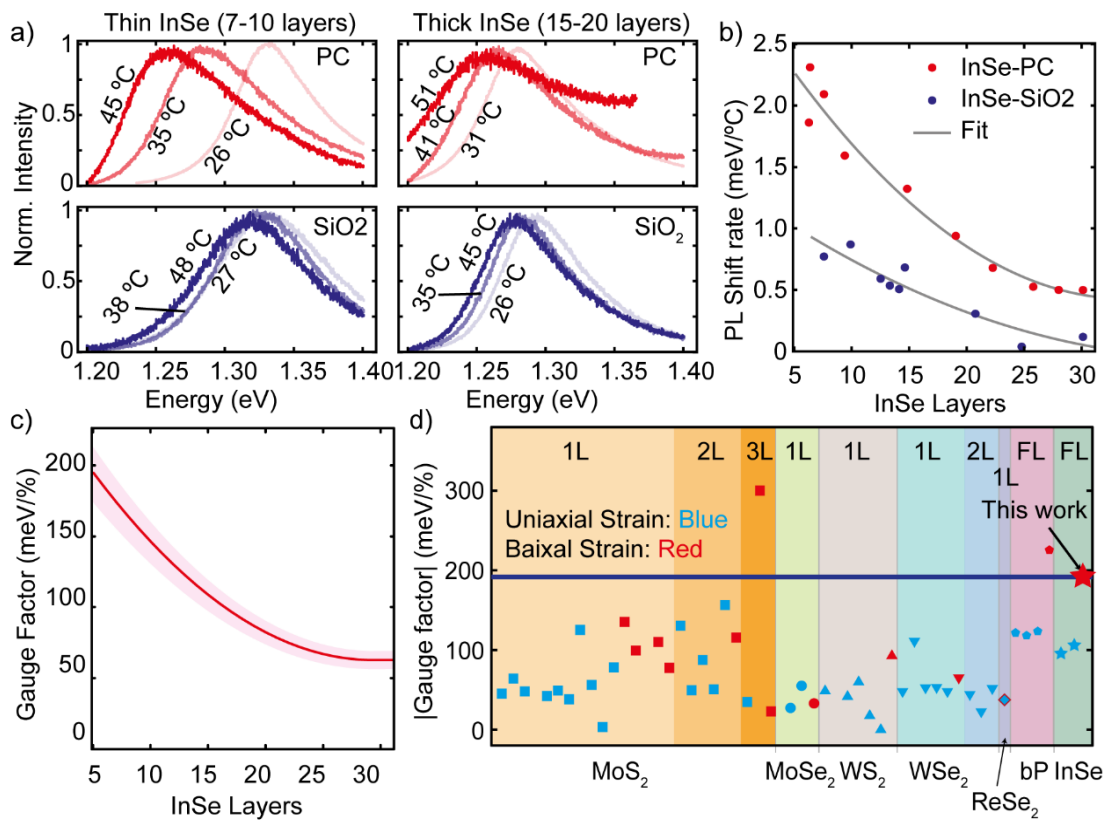

**Figure 3**. **Gauge factor of biaxial strained InSe.** a) Photoluminescence spectra of thin (7-10 layers, left) and thick (15-20 layers, right) InSe flakes deposited on PC substrate (top) and on 280 nm SiO$_2$/Si substrate (bottom) recorded as a function of temperature (from RT to ~ 50°C). b) PL energy shift rate versus thickness of InSe flakes deposited on PC substrate (red) and on 280 nm SiO$_2$/Si (blue) substrate. The solid lines represent the best fit for each dataset to a second order polynomial function. c) Calculated gauge factor of biaxial strained InSe flakes as a function of layer numbers. d) The different values of the strain tunability gauge factors of 2D semiconducting flakes reported in the literature with various approaches are compared with the value of biaxial strained InSe obtained in this work (~200 meV/%).

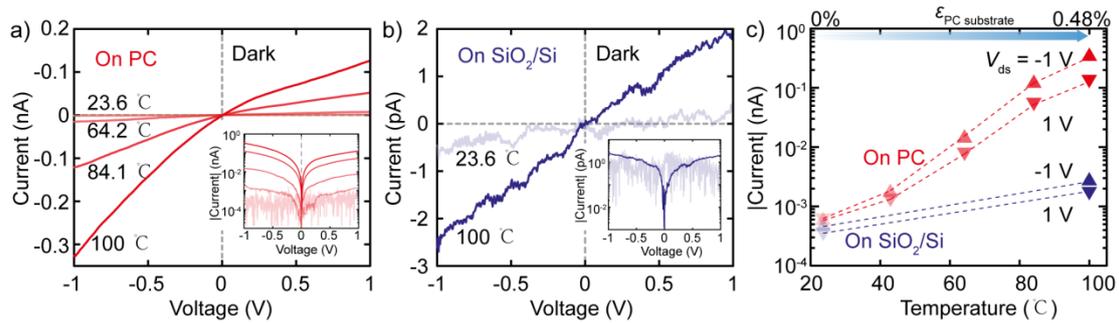

**Figure 4. Temperature-dependent dark current-voltage (*I-V*) characteristics of InSe device fabricated on PC and on 280 nm SiO$_2$/Si substrate.** a, b) *I-V* curves of Au-InSe-Au device on PC substrate (a) and on 280 nm SiO$_2$/Si substrate (b) recorded in dark conditions as a function of temperature (from ~23°C to ~100°C) in linear scale and semi-logarithmic scale (insets). c) Temperature-dependency of absolute values of the current flow through Au-InSe-Au devices at 1 V and -1 V on PC (red) and on 280 nm SiO$_2$/Si (blue) substrate. The top axis in (c) indicates the biaxial strain induced by the thermal expansion of the PC substrate.

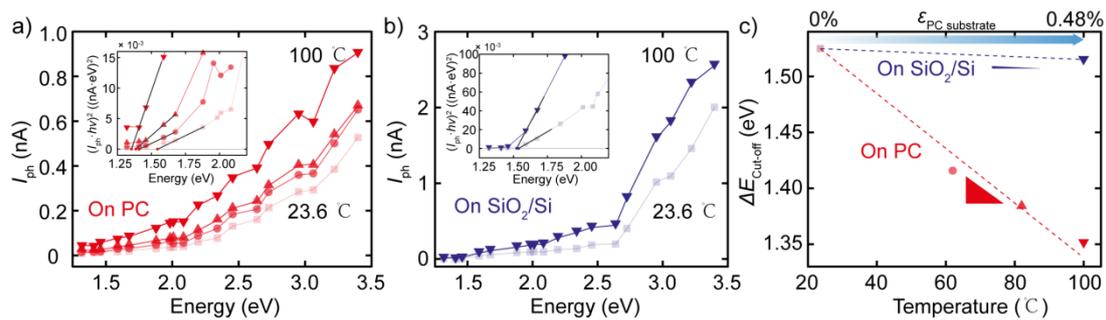

**Figure 5. Temperature-dependence of photocurrent spectra of InSe devices on PC and on 280 nm SiO$_2$/Si substrate.** a, b) Photocurrent ($I_{ph}$) *versus* illumination photon energy spectra recorded under a fixed illumination power intensity (35.4 mW/cm$^2$) at 1 V as a function of temperature (from ~23°C to ~100°C) of Au-InSe-Au device on PC (a) and on 280 nm SiO$_2$/Si substrate (b). Insets: Tauc plots: $(I_{ph} \cdot h\nu)^2$ *vs.* photon energy. c) Temperature-dependence of band gap values extracted from the Tauc plots as a function of temperature based on the Au-InSe-Au devices on PC (red) and on 280 nm SiO$_2$/Si (blue) substrate. The top axis in (c) indicates the biaxial strain induced by the thermal expansion of the PC substrate.

**Supporting Information:**

# Giant Piezoresistive Effect and Strong Band Gap Tunability in Ultrathin InSe upon Biaxial Strain

*Qinghua Zhao, Tao Wang\*, Riccardo Frisenda\*, Andres Castellanos-Gomez\**

Q. Zhao, Prof. T. Wang
State Key Laboratory of Solidification Processing, Northwestern Polytechnical University, Xi'an, 710072, P. R. China
Key Laboratory of Radiation Detection Materials and Devices, Ministry of Industry and Information Technology, Xi'an, 710072, P. R. China
E-mail: taowang@nwpu.edu.cn

Q. Zhao, Dr. R. Frisenda, Dr. A. Castellanos-Gomez
Materials Science Factory. Instituto de Ciencia de Materiales de Madrid (ICMM-CSIC), Madrid, E-28049, Spain.

E-mail: riccardo.frisenda@csic.es; andres.castellanos@csic.es

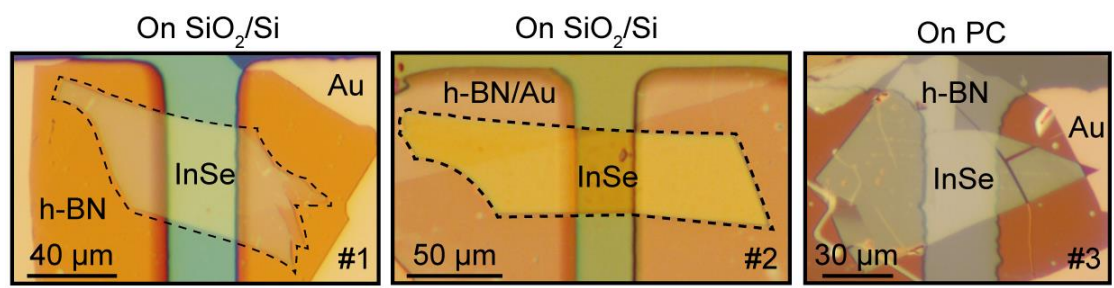

**Figure S1.** Three Au-InSe-Au devices fabricated on 280 nm SiO$_2$/Si (#1 and #2) and PC (#3) substrate.

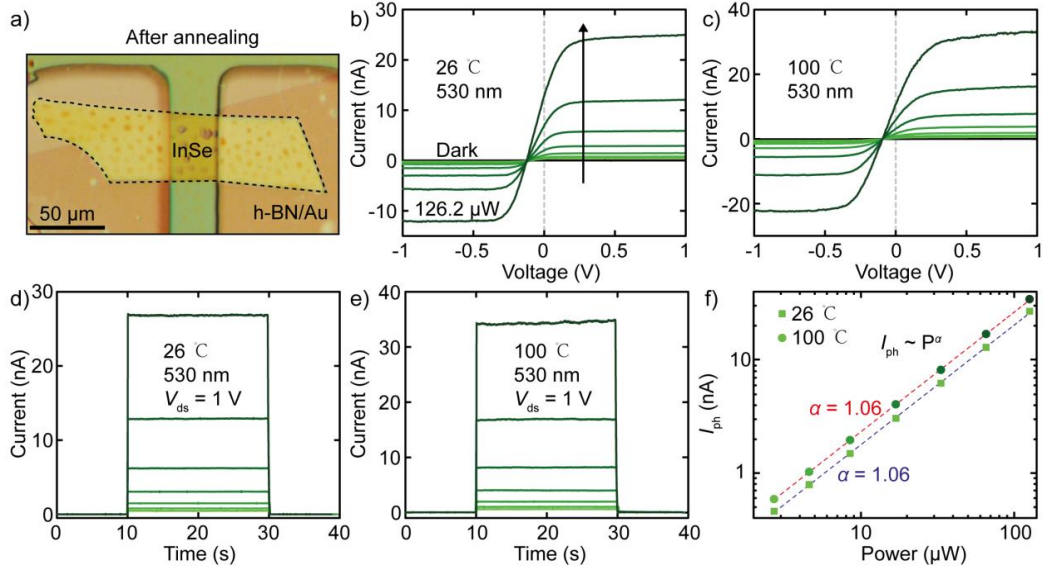

**Figure S2.** Optoelectronic characterization of device #2 after *in situ* annealing. a) The optical picture of device #2 after annealing. b, c) *I-V* curves recorded in dark condition and as a function of 530 nm illumination power at 26 °C and 100 °C. d, e) *I-t* curves at 1 V recorded as a function of 530 nm illumination power at 26 °C and 100 °C. f) Photocurrent value in the device at 1 V at 26 °C (square) and 100 °C (circle) *versus* illumination power plot in log-log scale.

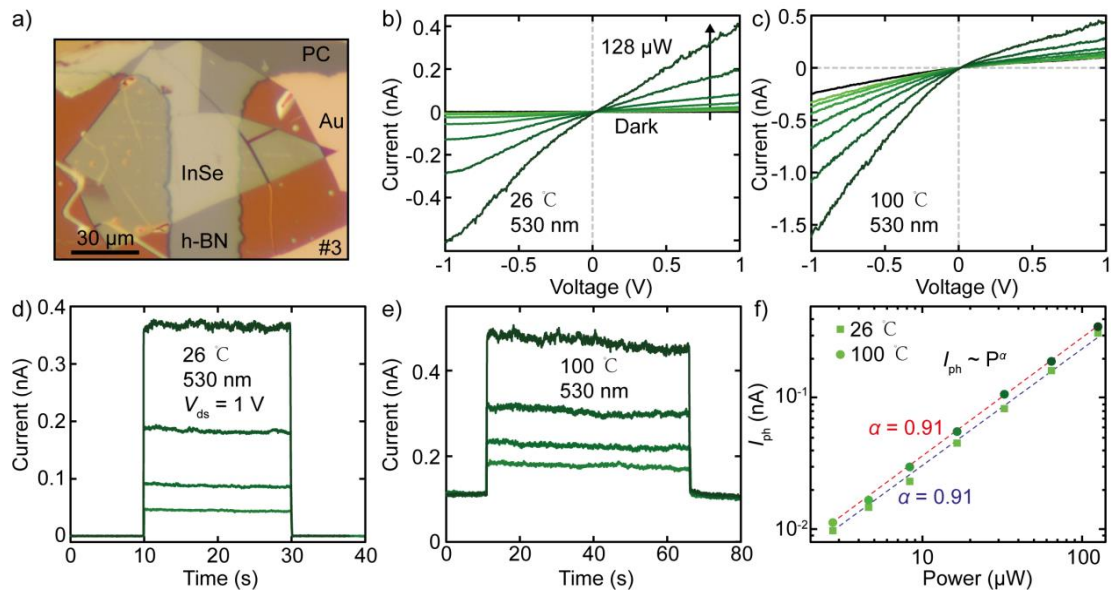

**Figure S3.** Optoelectronic characterization of device #3 after *in situ* annealing. a) The optical picture of device #3. b, c) *I-V* curves recorded in dark condition and as a function of 530 nm illumination power at 26 °C and 100 °C. d, e) *I-t* curves at 1 V recorded as a function of 530 nm illumination power at 26 °C and 100 °C. f) Photocurrent value in the device at 1 V at 26 °C (square) and 100 °C (circle) *versus* illumination power plot in log-log scale.

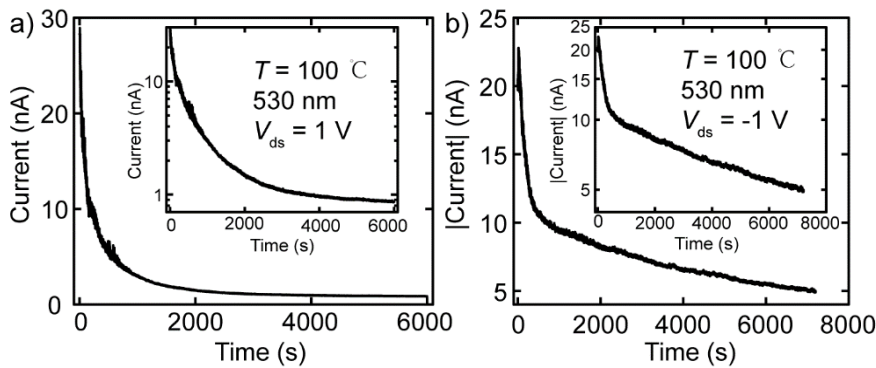

**Figure S4.** The current evolution during the *in situ* annealing. a, b) The current flowing through the device #1 at 1 V (a) and device #3 at -1 V (b) as a function of *in situ* annealing time in linear and semi-logarithmic (inset) scale.

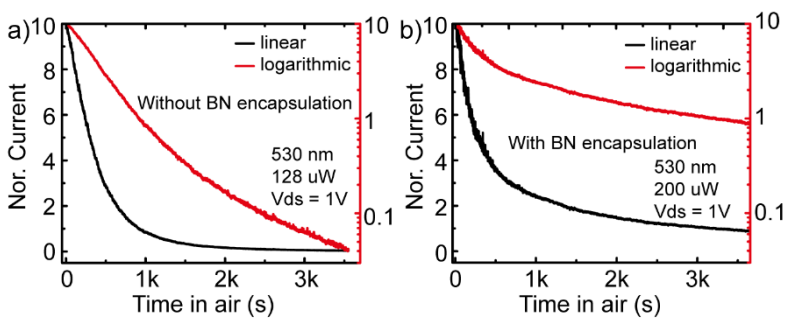

**Figure S5.** The current evolution during the *in situ* annealing of InSe devices without (a) and with (b) h-BN encapsulation.

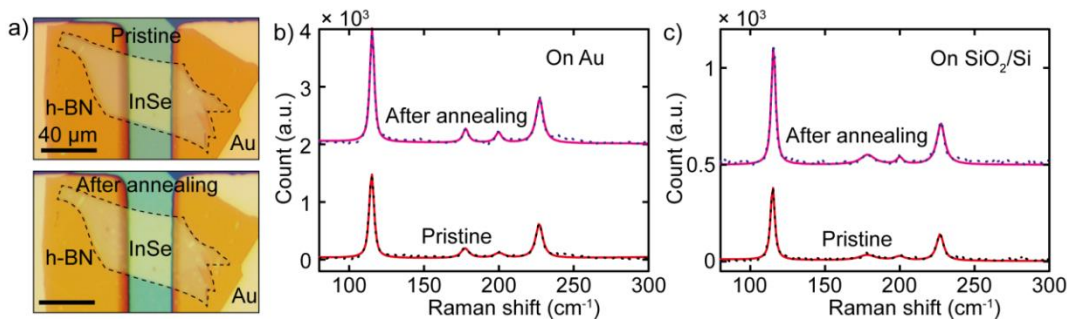

**Figure S6.** Structural stability before and after *in situ* annealing of Au-InSe-Au device. a) The optical picture of a Au-InSe-Au devices fabricated on $SiO_2/Si$ substrate before and after annealing. b, c) Raman Spectra of InSe flake on Au (b) and on $SiO_2/Si$ (c) recorded in pristine state and after annealing at room temperature.

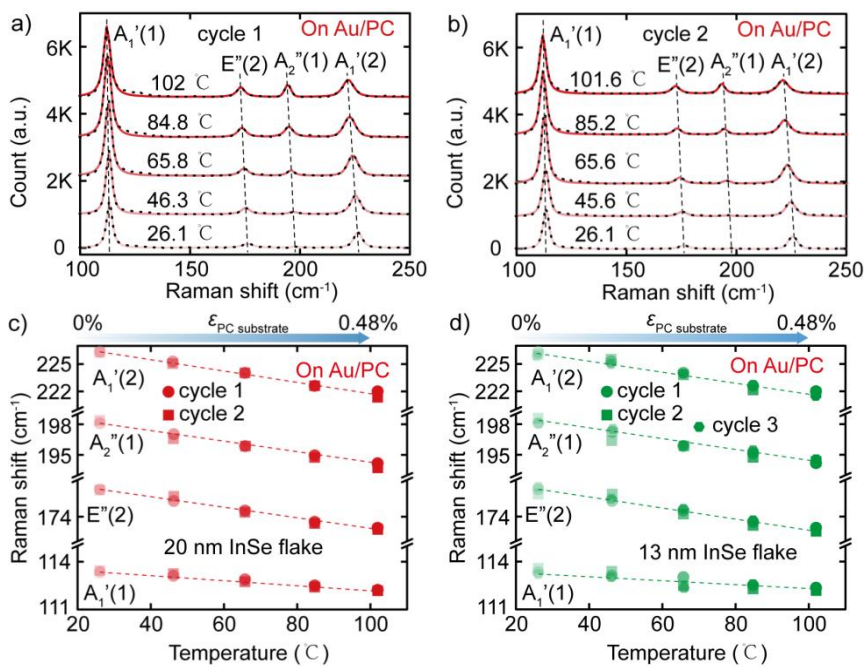

**Figure S7.** Temperature-dependent Raman Spectroscopy of thin InSe flakes recorded during different heating cycles. a, b) Raman spectra of thin InSe (~20 nm) recorded on Au/PC substrates during heating cycle 1 (a) and cycle 2 (b) with 50× objective as a function of temperature (from ~26°C to ~ 100°C). c-d) Temperature-dependency of four Raman active modes ($A_1'(1)$, $E''(2)$, $A_2''(1)$ and $A_1'(1)$) of 20 nm InSe (c) and 13 nm InSe (d) on PC substrate during different heating cycles. The top axis indicates the biaxial strain induced by the thermal expansion of the PC substrate.

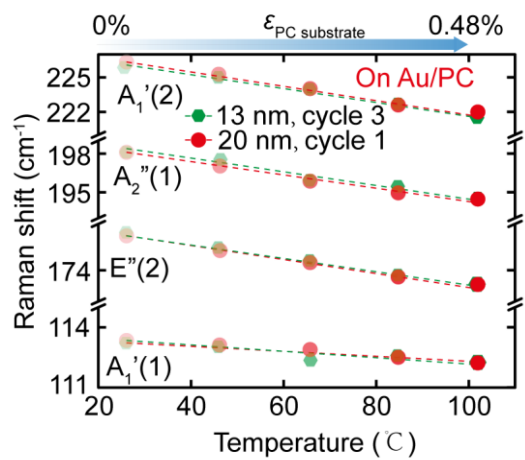

**Figure S8.** Temperature-dependency of four Raman active modes ($A_1'(1)$, $E''(2)$, $A_2''(1)$ and $A_1'(1)$) of two thin InSe flakes with the thicknesses of 13 nm (green) and 20 nm (red) on PC substrate. The top axis in (c) indicates the biaxial strain induced by the thermal expansion of the PC substrate.

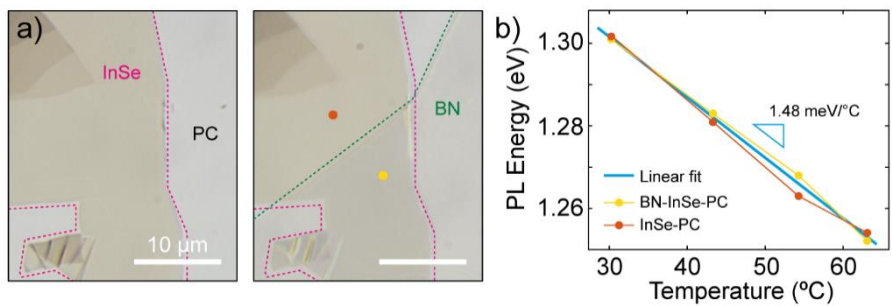

**Figure S9.** The effects of h-BN encapsulation on band gap tunability of biaxial strained InSe. (a) A thin InSe flake (~13 layers) deposited on PC substrate before (left) and after (right) being half encapsulated with h-BN. (b) The PL energy as a function of temperature of the InSe flake recorded at the two positions indicated by the colored circles in panel (a) corresponding to InSe-PC and BN-InSe-PC. The blue line is a linear fit to the data with slope 1.48 meV/°C.

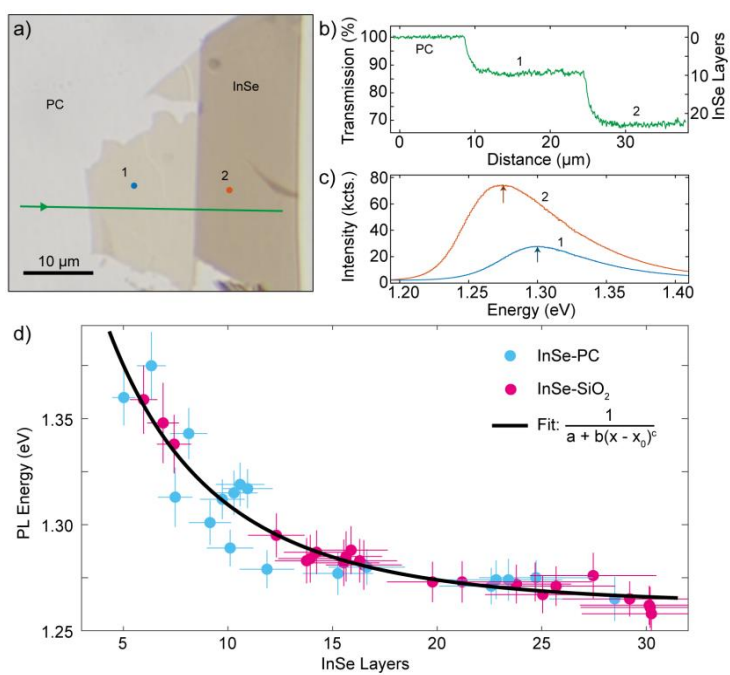

**Figure S10.** The PL energy versus InSe flake thickness. (a) Optical picture of a thin InSe flake with two different regions (marked by 1 and 2) deposited on PC substrate and recorded under transmission mode. b) Total transmission of the blue channel of the picture in panel (a) taken along the green line, with the PC having unitary transmission. The right axis indicates the corresponding estimated number of layers of the InSe flake. c) Photoluminescence spectra of regions 1 and 2 recorded at the locations marked by the red and blue circles in panel (a). d) Center of the PL peak as a function of InSe flake thickness extracted from 42 different InSe flakes deposited on PC and SiO$_2$. The black line is the best fit to the function shown in the legend with parameters: a=0.79, b=-1.3 · 10$^7$ c=-5.5, x$_0$=-26.

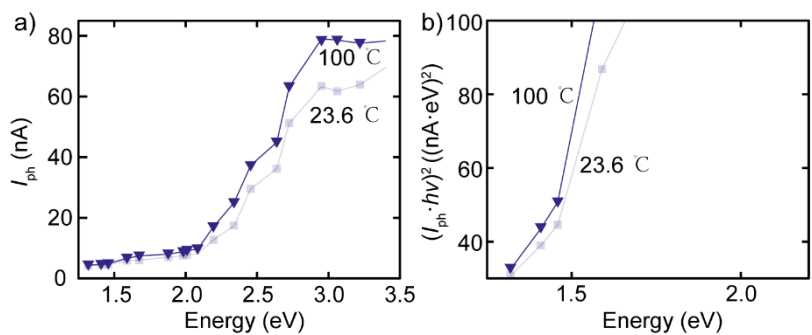

**Figure S11.** Temperature-dependency of photocurrent spectra (a) and Tauc plot (b) of device #2.